\documentclass[10pt]{extarticle}
\usepackage[a4paper, total={6in, 9in}]{geometry}
\usepackage[font=scriptsize]{caption}
\usepackage[english]{babel}
\usepackage{graphicx}
\usepackage{amsmath}
\usepackage{caption}
\usepackage{subcaption}
\usepackage{rotating}
\usepackage{url}
\usepackage{hyperref}
\usepackage{fancyhdr}
\usepackage{float}
\usepackage{cite}
\usepackage{authblk}
\usepackage{ulem}
\usepackage[usenames, dvipsnames]{color}

\begin{document}
\renewcommand\Authfont{\small}
\renewcommand\Affilfont{\itshape\footnotesize}
\title{Laser induced THz emission from femtosecond photocurrents in Co/ZnO/Pt and Co/Cu/Pt multilayers}
\author[1]{G. Li\footnote{Email: Qiao.Li@science.ru.nl}}
\author[1]{R.V. Mikhaylovskiy}
\author[2]{K.A. Grishunin}
\author[3]{J.D. Costa}
\author[1]{Th. Rasing}
\author[1,2]{A.V. Kimel}
\affil[1]{Radboud University, Heyendaalseweg 135, 6525 AJ, Nijmegen, The Netherlands.}
\affil[2]{Moscow Technological University (MIREA), Vernadsky Ave. 78, 119454, Moscow, Russia.}
\affil[3]{International Iberian Nanotechnology Laboratory (INL), Braga, Portugal}
\date{\today}
\maketitle

\begin{abstract}
The ultrashort laser excitation of Co/Pt magnetic heterostructures can effectively generate spin and charge currents at the interfaces between magnetic and nonmagnetic layers. The direction of these photocurrents can be controlled by the helicity of the circularly polarized laser light and an external magnetic field. Here, we employ THz time-domain spectroscopy to investigate further the role of interfaces in these photo-galvanic phenomena. In particular, the effects of either Cu or ZnO interlayers on the photocurrents in Co/X/Pt (X = Cu, ZnO) have been studied by varying the thickness of the interlayers up to 5 nm. The results are discussed in terms of spin-diffusion phenomena and interfacial spin-orbit torque.
\end{abstract}

\section{Introduction}
The idea to use not only the charge but also the spin of electrons in the operation of electronic devices has led to the development of spintronics, causing a revolution on how information is stored and processed \cite{jungwirth2012,otani2017}. A new challenge is to develop ultrafast spintronics. The use of femtosecond laser pulses and photo-galvanic phenomena is a promising approach for ultrafast generation and control of photocurrents \cite{huisman2016_2,walowski2016}. Two mechanisms of such generation have been demonstrated in Co/Pt bilayers, which is one of the most popular model systems in spintronic research \cite{kimel2005,nemec2012,hoffmann2013,sinova2015}.

The first mechanism is based on ultrafast laser-induced demagnetization of Co grown such that it has in-plane magnetization \cite{tserkovnyak2002,ando2011,battiato2012}. The demagnetization of the Co layer results in a spin current injection into the Pt layer. Then, due to the inverse spin-Hall effect, originating from the spin-orbit interaction in the Pt-layer, the spin current is converted into a charge current \cite{kampfrath2013,seifert2016,huisman2017}. The direction of the charge current is perpendicular to both the magnetization and the normal to the Co/Pt interface. The polarity of the current can be changed by changing the polarity of the magnetization. The current does not depend on the polarization of light.

The second mechanism requires circularly polarized optical excitation. It is similar to the spin-galvanic effect reported first by Ganichev et al. for non-centrosymmetric semiconductors \cite{ganichev2002}. It was shown that electrical photocurrents can be induced at the interface of Co/Pt, where the direction of the current can be switched by changing the magnetization and the polarization of light \cite{huisman2016}. Broken inversion symmetry in combination with spin-orbit interaction at the interface are essential for a non-zero spin-orbit torque responsible for the reciprocal connection between the electrical current and the magnetization tilting \cite{manchon2015,garello2013,freimuth2014,freimuth2015,freimuth2016}. The optical tilting of the magnetization can happen either via the inverse Faraday effect or the optical spin transfer torque, respectively \cite{choi2017,kimel2005,nemec2012}. It was suggested that in the case of Co/Pt with the Co-layer magnetized in the sample plane, this torque produces an electrical current in the direction perpendicular to the Co/Pt interface and parallel to the magnetization \cite{huisman2016}.

Interfaces responsible for the inversion symmetry breaking must play an important role for both mechanisms of the control of femtosecond photocurrents in Co/Pt bilayers. In particular, these mechanisms must depend on the nature and thickness of the interlayer between Co and Pt. The goal of this work is to elucidate how metallic Cu and semiconducting ZnO interlayers affect the generation of photocurrents in Co/Pt heterostructures. With the help of THz emission spectroscopy we find that the helicity dependent laser induced THz emission to be more sensitive to the interface than the helicity independent one. This helicity dependent emission can be explained in terms of spin-galvanic effects, for which the intermixing of spin-polarized electrons of Co and Pt is crucial.

\section{Experimental Method}
\subsection{General Principles}
The generation of THz electromagnetic radiation by ultrashort pulses of electric currents can be understood on the basis of the Maxwell equations:
\begin{align}
\label{eq:maxwell}
\nabla \times {\bf E} &= -\frac{\partial {\bf B}}{\partial t}, \\
\nabla \times {\bf H} &= \frac{\partial {\bf D}}{\partial t} + {\bf J}.
\end{align}
Here {\bf J} is the free charge current, {\bf D} is the electric displacement field defined as  ${\bf D} = \epsilon_{0}{\bf E} + {\bf P}$, {\bf P} is the polarization, {\bf E} is the strength of the electric field. {\bf H} is the strength of the magnetic field
defined as ${\bf H} = \frac{1}{\mu_{0}}{\bf B} - {\bf M}$, {\bf M} is the magnetization, and {\bf B} is the magnetic flux density. $\epsilon_{0}$ and $\mu_{0}$ are the dielectric and magnetic constants, respectively. In the case of no static charges the wave equation in free space in terms of the electric field can be written as
\begin{equation}
\label{eq:wave_eq}
\nabla^{2} {\bf E} - \mu_{0}\epsilon_{0}\frac{\partial^{2} {\bf E}}{\partial t^{2}} = \mu_{0} \frac{\partial}{\partial t}\left( {\bf J} + \nabla \times {\bf M} \right).
\end{equation}
According to Eq.\ref{eq:wave_eq}, emitted THz electric fields contain information on the ultrafast dynamics of charge currents and magnetization \cite{huisman2015}. In the case of the studied Co/X/Pt (X = Cu or X = ZnO) heterostructures the emitted THz electric field was shown to be mainly induced by ultrafast charge current dynamics \cite{huisman2016}. This is confirmed by our experiments.

\subsection{THz Emission Spectroscopy}
To excite the studied samples we employed 40 fs circularly polarized laser pulses at a repetition rate of 500 Hz with a central wavelength at 800 nm and a fluence of 0.6 $\mathrm{mJ}$ $\mathrm{cm^{-2}}$. The helicity of the circularly polarized light was varied from left to right handedness using a quarter wave-plate. An external bias magnetic field up to 0.1 Tesla was applied in the sample plane.

The emitted THz radiation was focused by a pair of gold coated parabolic mirrors onto a 1 mm thick ZnTe crystal. When the THz radiation passes through the ZnTe crystal it induces a change in the birefringence proportional to the amplitude of the electric field. This phenomenon is called the Pockels effect. The linear birefringence induced in this crystal by the emitted THz electric field pulse was probed by another ultrashort laser pulse at the wavelength of 800 nm. The THz-induced birefringence can be seen as the ellipticity of the polarization of the laser pulse acquired after interaction with the ZnTe crystal. To measure the ellipticity and thus to quantify the strength of the emitted THz electric field, we employed a balance detector.

In the chosen coordinate system the pump beam and the emitted THz radiation were directed along the \^{z}-axis, which was also the normal to the sample (see Fig.\ref{fig:graphs_even}(a) and Fig.\ref{fig:graphs_odd}(a)). The magnetic field was always applied along the \^{x}- axis either up ($\mathrm{M^{+}}$) or down ($\mathrm{M^{-}}$). The experiments were performed with left- ($\sigma^{+}$) and right- ($\sigma^{-}$) handed circularly polarized pump pulses. In order to probe different mechanisms of photocurrent generation in the experiments, the vertical $E_{x}$ and the horizontal $E_{y}$ components of the emitted THz electric field were measured with the help of a wire-grid polarizer between the sample and the electro-optical crystal ZnTe.

In the following we will distinguish THz emission signals which depend on the polarization of the pump from the one which are not affected by the change of the pump polarization. The helicity dependent ($E_{x,odd}$) and independent ($E_{y,even}$) THz signal of the electric field radiation can be calculated from the experimental data in the following way:
\begin{align}
\label{eq:helicity_odd}
E_{x,odd}  = \frac{1}{4}(E_{x}^{\sigma^{+}M^{+}}-E_{x}^{\sigma^{-}M^{+}}-E_{x}^{\sigma^{+}M^{-}}+E_{x}^{\sigma^{-}M^{-}}), \\
\label{eq:helicity_even}
E_{y,even} = \frac{1}{4}(E_{y}^{\sigma^{+}M^{+}}+E_{y}^{\sigma^{-}M^{+}}-E_{y}^{\sigma^{+}M^{-}}-E_{y}^{\sigma^{-}M^{-}}).
\end{align}
Here $E_{x,odd}$ is the \^{x}-component of the helicity dependent THz signal and  $E_{y,even}$ is the \^{y}-component of the helicity independent THz signal. $E^{\sigma^{\pm}M^{\pm}}$ is the emitted THz electric field for the case when the sample is excited by a circularly polarized femtosecond pump pulse of the chosen helicity ($\sigma^{\pm}$). The polarity of the magnetization of the Co-layer ($M^{\pm}$) is oriented along the \^{x}-axis.

\subsection{Sample Fabrication} 
The studied multilayer samples were deposited in a multi-target UHV sputtering system with a base pressure of $5 \times 10^{-8}$ torr. For all samples the procedure starts with the fabrication of a 10 nm thick ferromagnetic layer of Co deposited on a 500 $\mu$m thick glass substrate. Then, either a layer of Cu or ZnO is deposited onto Co. The interlayer is then capped with a 2.6 nm thick Pt layer. Two additional samples were fabricated with the Pt layer directly on the Co layer. The deposition conditions were optimized for each individual material. For the deposition of Co, Cu, and Pt a DC current of 100 mA, 75mA, and 100 mA and an Ar flux of 7 sccm, 20 sccm, and 10 sccm was applied, respectively. For ZnO an Ar flow of 15 sccm was applied and a RF power of 50 W was needed due to charge accumulation. These conditions give depositions rates smaller than 1 $\mathrm{\AA}$ $\mathrm{s^{-1}}$, which allow for a precise control of the layer thickness. All samples were deposited on top of a magnetic pallet, a permanent magnet that induced a uniaxial in plane magnetic anisotropy on the magnetic films. X-ray reflectivity characterization of similar grown Co/Pt samples, as reported in Ref.\cite{huisman2016}, showed that the interface roughness between the Co and Pt layers is about 1 nm. The studied samples were grown following a similar procedure and thus also in this case a high roughness of the boundaries between the layers in the studied heterostructures is expected.

\section{Experimental Results}
\subsection{THz Emission From the Co/Pt Bilayers}
\begin{figure}
	\centering
	\includegraphics[width=1\linewidth]{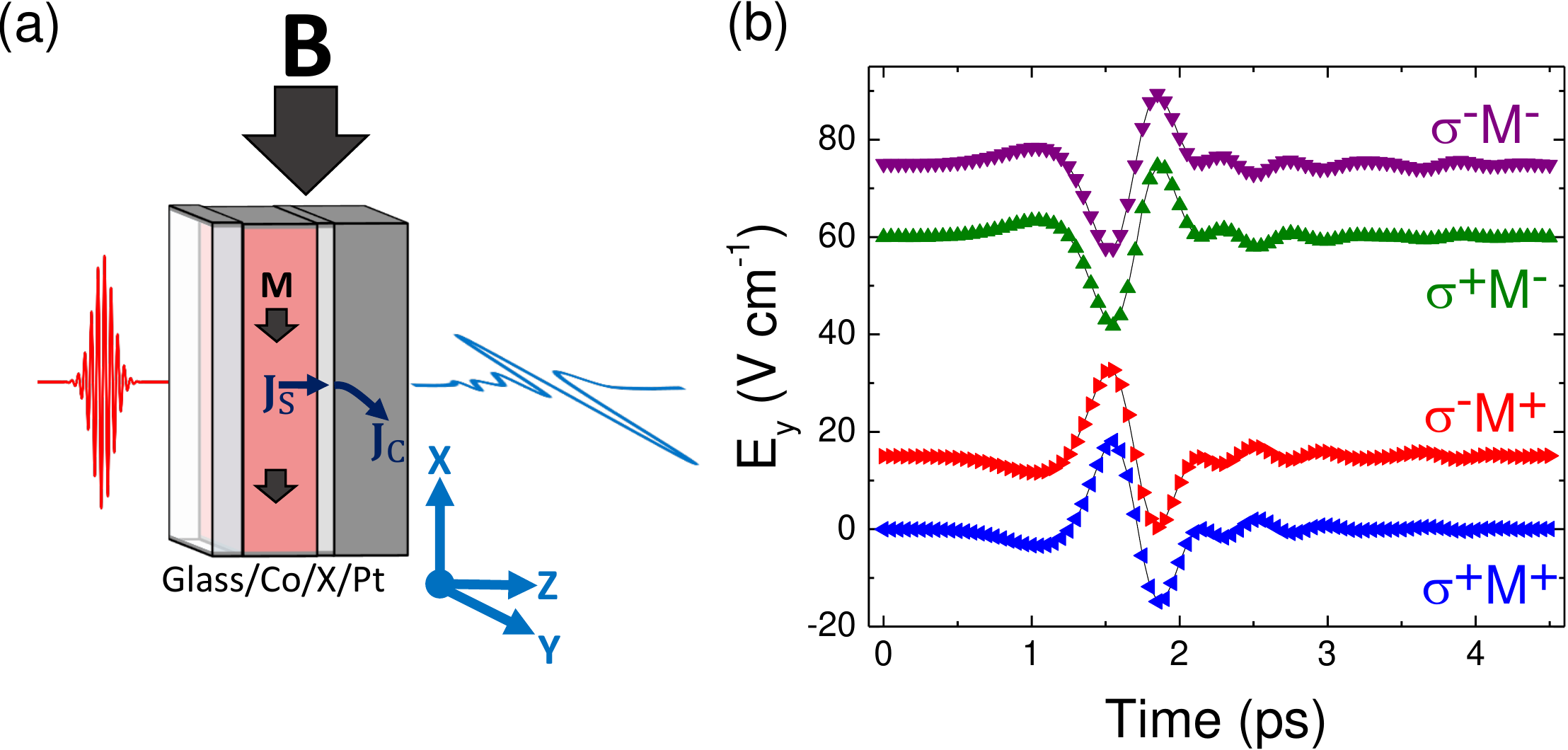}
	\caption{\label{fig:graphs_even}(a) The experimental geometry for the detection of the helicity independent THz emission in Co/X/Pt with X = ZnO or Cu. An ultrashort laser pulse induces a spin current from the ferromagnetic layer (Co) \cite{kampfrath2013}. The spin current (${\bf J}_{\mathrm{S}}$) flowing along the \^{z}-axis will then get converted into a charge current (${\bf J}_{\mathrm{C}}$) along the \^{y}-axis via the inverse spin-Hall effect. (b) The time traces of the electric field component $E_{y}$ measured for the Co/Pt bilayer for two polarities of the magnetization ($M^{+}$, $M^{-}$) and two helicities of light ($\sigma^{+}$, $\sigma^{-}$). No dependence on the helicity of the pump pulse is seen. The traces experience a sign change upon magnetization reversal in Co.}
\end{figure}
\begin{figure}
	\centering
	\includegraphics[width=1\linewidth]{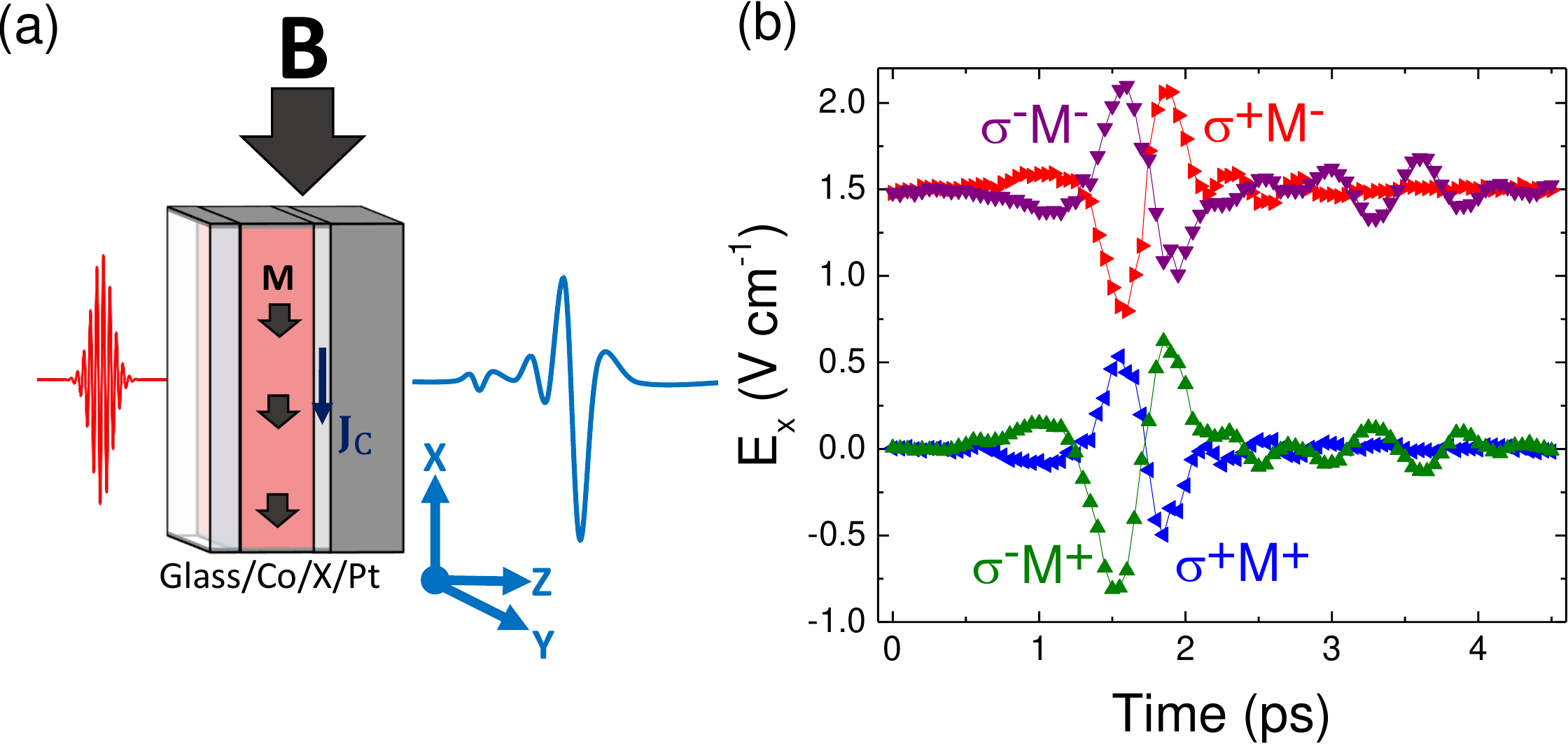}
	\caption{\label{fig:graphs_odd}(a) The experimental geometry for the detection of the helicity dependent THz emission in Co/X/Pt with X = ZnO or Cu. An ultrashort laser pulse induces an in-plane charge current (${\bf J}_{\mathrm{c}}$) at the interface \cite{huisman2016}. The direction of this charge current is dependent on the helicity of the pump pulse. (b) The time traces of the electric field component $E_{x}$ measured for the Co/Pt bilayer for two polarities of the magnetization ($M^{+}$, $M^{-}$) and two helicities of light ($\sigma^{+}$, $\sigma^{-}$). Clear dependence on the helicity of the pump pulse is seen. The traces experience also a sign change upon magnetization reversal in Co.}
\end{figure}
\begin{figure}
	\centering
	\includegraphics[width=1\linewidth]{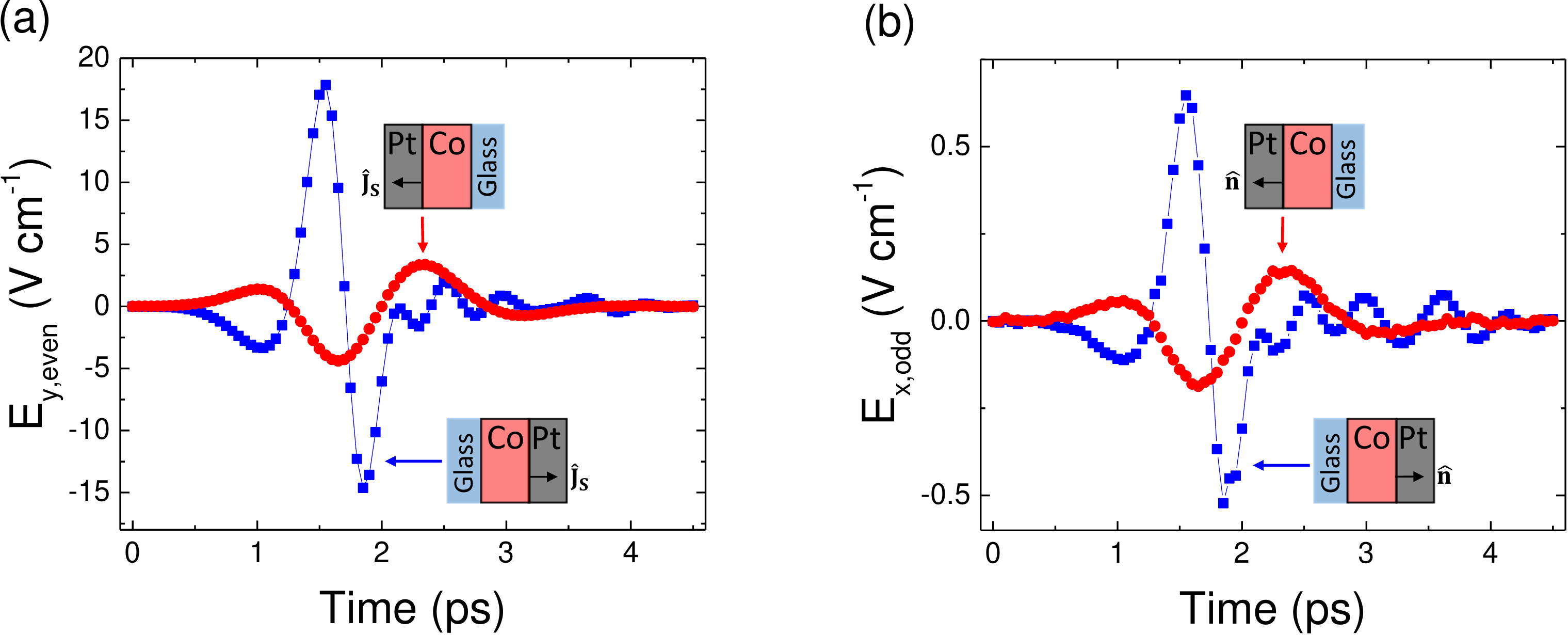}
	\caption{\label{fig:graphs_helicity}(a) Time-resolved traces of the helicity independent electric field $E_{y,even}$ measured for two orientations of the sample (Co/Pt, Pt/Co). The orientation defines the polarity of the vector corresponding to the spin current ${\bf J}_{\mathrm{S}}$ in the laboratory coordinate system. (b) Time-resolved traces of the helicity dependent electric field $E_{x,odd}$ measured for two orientations of the sample. The orientation now defines the polarity of the vector ${\bf n}$ which characterizes symmetry breaking at the Co/Pt interface and points from the Co to the Pt layer. $E_{x,odd}$ and $E_{y,even}$ are calculated from the raw data according to Eq.\ref{eq:helicity_odd} and \ref{eq:helicity_even}.}
\end{figure}
The geometry of the experiments and the results of the time-resolved measurements of the emitted THz electric field are shown in Fig.\ref{fig:graphs_even}, for the case when the polarization of the electric field is along the \^{y}-axis. The experiments were performed for two polarities of the magnetization and two helicities of the pumping light. It is seen that only a change of the magnetization results in a different outcome of the experiment and the measured dynamics is helicity independent. A spin current is generated from the Co layer due to the demagnetization by the femtosecond pump laser pulse. This current flows into the Pt layer where, due to a strong spin-orbit interaction, the electrons are deviated in the direction perpendicular to their momentum and spin. This mechanism is known as the inverse spin-Hall effect \cite{hoffmann2013,sinova2015}. To confirm this, we compared the detected THz signals from two configurations: the first configuration is where the pump is incident on Co/Pt, the second is where the pump is incident on Pt/Co. Further in this article we will use notations like Co/Cu/Pt or Co/ZnO/Pt meaning that light is incident first on the glass substrate and the Co layer. Notations Pt/Cu/Co or Pt/ZnO/Co mean that light is incident from the opposite side hitting the Pt layer first. Figure \ref{fig:graphs_helicity}(a) reveals the asymmetric THz field traces for the two directions of the spin current. The weaker THz signal and longer pulse is because the emitted THz field has to travel through the glass substrate in the Pt/Co pump configuration and suffers dispersion and absorption.

Measuring the polarization of the detected THz signal allows us to reveal different photocurrents (see Fig.\ref{fig:graphs_odd}). It is seen that the dynamics is sensitive to both the magnetization and the helicity of the pump light. A possible origin of the  helicity dependent THz signal ($E_{x,odd}$) is the spin-galvanic effect at the interface, where it shows Rashba symmetry and broken space inversion symmetry \cite{ganichev2002,freimuth2016,huisman2016}. The origin of the helicity independent THz signal ($E_{y,even}$) is understood as spin polarized electrons coming from the demagnetized Co layer \cite{battiato2012,kampfrath2013}. The asymmetric THz signal seen in Fig.\ref{fig:graphs_helicity}(b) indicates the significance of the broken space inversion symmetry for the photocurrent. Moreover, the asymmetry in both the electric field polarizations means that the emitted THz electric field must be of electric dipole origin. It shows that the dynamics of the $E_{x}$ and $E_{y}$ components correspond to the THz radiation emitted by the helicity dependent and independent photocurrents, respectively. We emphasize that these two mechanisms generate simultaneously a THz electric field radiation with perpendicular polarization with respect to each other. The spin to charge conversion via the inverse spin-Hall effect or charge current excitation via spin-orbit torque can therefore be studied independently in one experiment, detecting THz light of two mutually orthogonal polarizations. Evidently, interfaces play a significant role for the helicity dependent phenomena and can be important for the helicity independent phenomena. To study the significance of the Co/Pt interface we introduced two types of interlayers (ZnO and Cu) and studied the generated photocurrent as a function of the interlayer thickness.

\subsection{The Role of the ZnO Interlayer}
\begin{figure}
	\begin{subfigure}{.49\linewidth}
		\includegraphics[width=\linewidth]{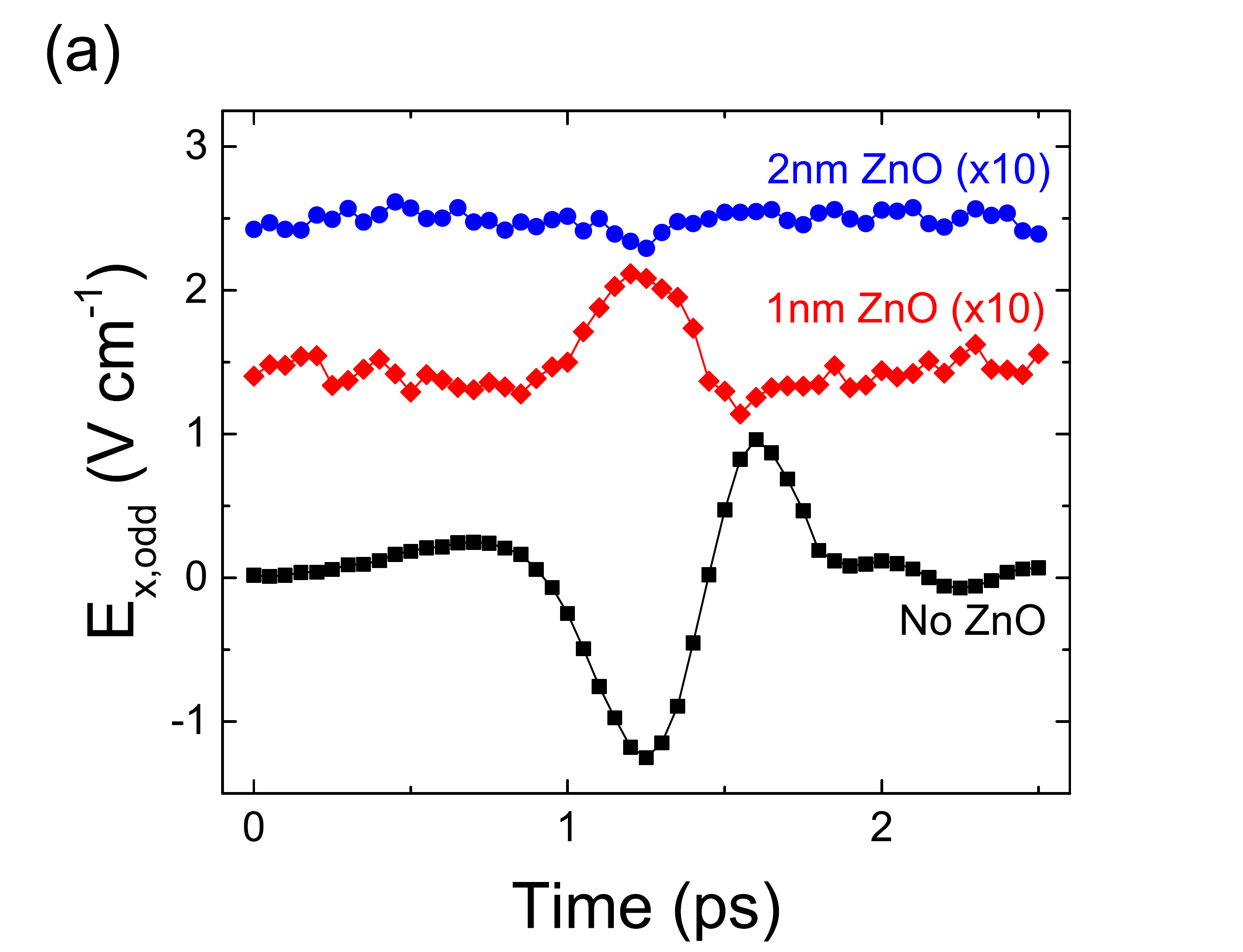}
	\end{subfigure}
	\begin{subfigure}{.49\linewidth}
		\includegraphics[width=\linewidth]{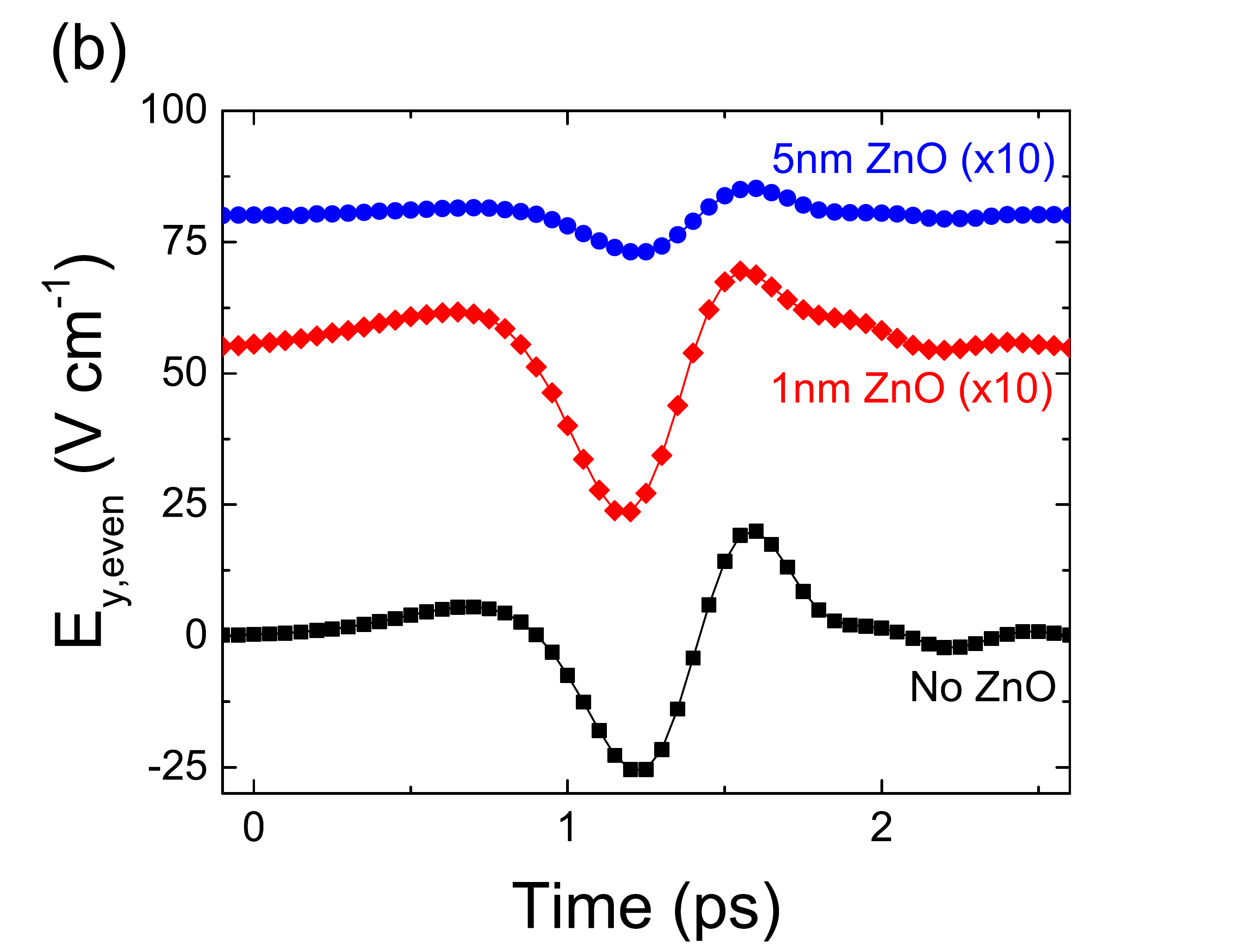}
	\end{subfigure}
	\caption{\label{fig:ZnO_odd-even} Time traces of the (a) helicity dependent ($E_{x,odd}$) and (b) helicity independent ($E_{y,even}$) component for different ZnO interlayer thicknesses. The pump laser pulse has a central wavelength of 800 nm and the experiments were performed for the Co/ZnO/Pt orientation of the structures.}
\end{figure}
\begin{figure}
	\begin{subfigure}{.49\linewidth}
		\includegraphics[width=\linewidth]{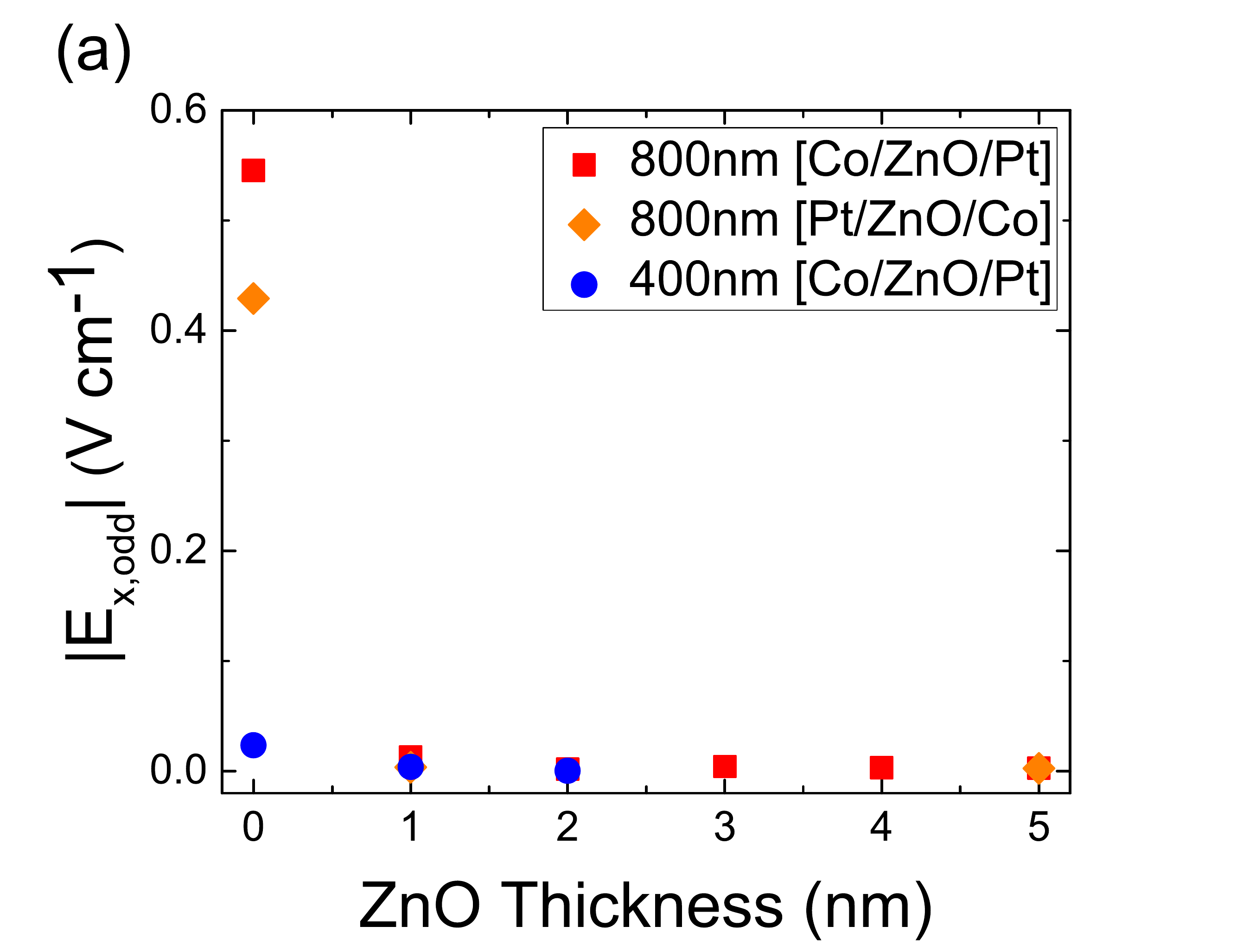}
	\end{subfigure}
	\begin{subfigure}{.49\linewidth}
		\includegraphics[width=\linewidth]{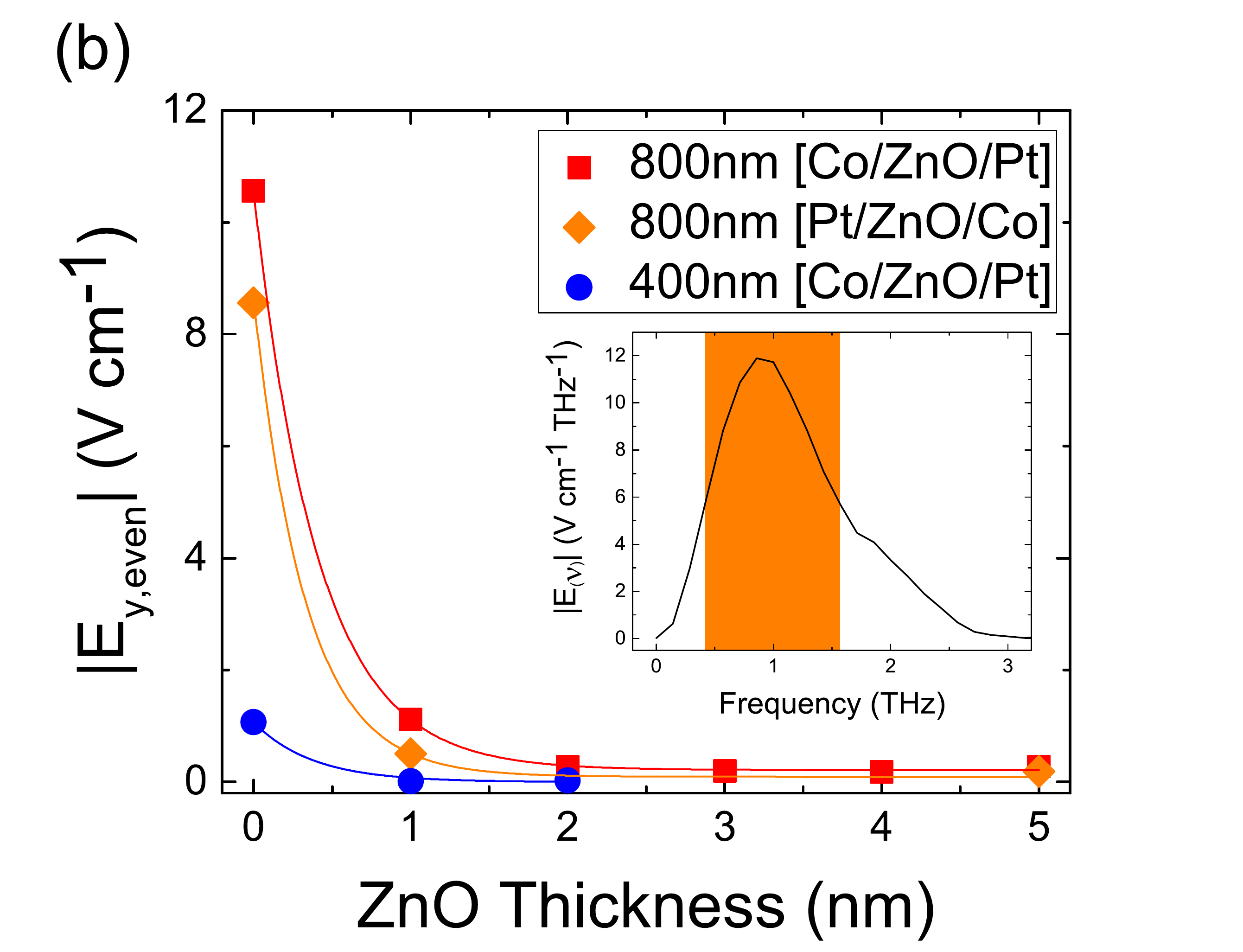}
	\end{subfigure}
	\caption{\label{fig:ZnO_spectrum-results} The integral strength of the THz electric field as a function of the ZnO interlayer is obtained by integrating the amplitude of the Fourier spectrum $|E|$ over the frequency range between two cut-off frequencies corresponding to the half-maximum. (a) The helicity dependent $|E_{x,odd}|$ and (b) the helicity independent $|E_{y,even}|$ integral THz field. Squares (red) and rhombuses (orange) correspond to the THz emission excited by a pump pulse with a central wavelength of 800 nm and a fluence of 0.6 $\mathrm{mJ}$ $\mathrm{cm^{-2}}$ incident on Co/ZnO/Pt and Pt/ZnO/Co, respectively. The circles (blue) show the THz electric field corresponding to a pump at a central wavelength of 400 nm and fluence of 35 $\mathrm{\mu J}$ $\mathrm{cm^{-2}}$ incident on Co/ZnO/Pt. The solid lines are the fitted exponential decay functions. The inset on panel (b) shows an exemplary spectrum of the measured time trace signal highlighting the range for the integration (orange).}
\end{figure}
Aiming to reveal the effects of a ZnO interlayer on the generation of photocurrents and subsequent THz radiation, three different types of experiments were performed. In the first two experiments the THz emission was studied as a function of the interlayer thickness for two orientations of the sample, Co/ZnO/Pt and Pt/ZnO/Co, respectively. A pump pulse with a fluence of 0.6 $\mathrm{mJ}$ $\mathrm{cm^{-2}}$ and a central wavelength of 800 nm was used for the excitation. The third type of experiment was performed with a pump fluence of 35 $\mathrm{\mu J}$ $\mathrm{cm^{-2}}$ with a central wavelength of 400 nm. As ZnO is a semiconductor with a bandgap of 3.4 eV, it is interesting to check if a change in the pump photon energy from 1.5 eV to 3.1 eV will affect the studied thickness dependencies. These experiments were successful only for the Co/ZnO/Pt orientation of the sample. Similar experiments for the case of Pt/ZnO/Co orientation of the sample in principle give smaller THz signals due to absorption of the THz light in the glass substrate. Due to the absorption in the substrate, the THz signal generated by a pump fluence of 35 $\mathrm{\mu J}$ $\mathrm{cm^{-2}}$ and a central wavelength of 400 nm for the orientation Pt/ZnO/Co was below the sensitivity of our experimental setup. Moreover, for the Co/ZnO/Pt orientation the THz signal was only visible for a ZnO interlayer thickness up to 2 nm.

We start with the first two experiments of the pump pulse at a central wavelength of 800 nm. With the introduction of the ZnO interlayer a strong suppression of the THz signal is observed (see Fig.\ref{fig:ZnO_odd-even}). In Fig.\ref{fig:ZnO_odd-even}(a) the helicity dependent component $E_{x,odd}$ is fully suppressed after separating the Co and Pt layers by just 2 nm of ZnO while for 1 nm ZnO the helcity dependent THz signal is small. As previously mentioned the interface roughness is assumed to be 1 nm, which might explain the small THz signal at 1 nm ZnO. Due to the high roughness the formation of Co-islands during the growth is very likely where small regions of Co still in contact with the Pt layer may exist. Increasing the thickness to up to 2 nm results in a complete suppression of $E_{x,odd}$. This indicates the importance of a direct electric contact between the Co and Pt layers for the generation of the helicity dependent photocurrent. The helicity independent component $E_{y,even}$ also shows a strong decay of the amplitude, but it is still clearly visible up to a thickness of 5 nm (see Fig.\ref{fig:ZnO_odd-even}(b)).

To estimate the THz electric field strength we calculated the Fourier spectra of $E_{y,even}$ and $E_{x,odd}$ time traces, the amplitude of every spectrum peaks roughly around 1.5 THz. Absorption of ZnTe suppresses the THz signal above 3 THz, resulting in a strong suppression of the sensitivity of the THz detection above this frequency. At low frequencies the sensitivity is suppressed due to the fact that the wavelength of the radiation becomes comparable with characteristic sizes of our spectrometer\cite{gallot1999}. The amplitude of the THz signal is generally lower for the Pt/X/Co configuration due to the absorption of the THz emission in the glass substrate. To quantify the strength of the spectrally broad THz emision we introduce the integral THz electric field. To calculate this quantity we first define the sensitivity range by setting two cut-off frequencies at both sides of the peak (see the orange region in the inset of Fig.\ref{fig:ZnO_spectrum-results}(b)). These cut-off frequencies are defined as those corresponding to 50\% of the peak amplitude. By integrating over this frequency range we obtain the integrated field strength $|E|$ of the helicity independent and dependent THz signal.

Figure \ref{fig:ZnO_spectrum-results} shows the integrated helicity dependent and independent THz field strengths. The uncertainty in the interlayer thickness depends on the fabrication process, we assume it to be 1 nm corresponding to the interface roughness. The integrated field strength value for each interlayer thickness was averaged over a minimum of 3 measurements and the margin of error extraced from the measurements was less than 5\%. We therefore do not show the error bars for the sake of clarity. Figure \ref{fig:ZnO_spectrum-results}(a) confirms that for different pump incidences and pump wavelengths, the helicity dependent THz signal $|E_{x,odd}|$ is almost fully suppressed after separation of the Co and Pt layers. On the other hand, the helicity independent THz signal $|E_{y,even}|$ reveals a strong decay with respect to the ZnO thickness (see Fig.\ref{fig:ZnO_spectrum-results}(b)). The decay of $|E_{y,even}|$ is then phenomenologically found to be exponentially decreasing with interlayer thickness as
\begin{equation}
\label{eq:fit}
|E| = |E_{0}| e^{-\frac{z}{l}} + |E_{bg}|,
\end{equation}
where $|E|$ is the integrated field intensity of the THz electric field representing the current density, $l$ is the diffusion length, and $|E_{bg}|$ is an offset due to the background noise. The corresponding fits are shown in Fig.\ref{fig:ZnO_spectrum-results}(b) for Co/ZnO/Pt.

\subsection{The Role of the Cu Interlayer}
\begin{figure}
	\begin{subfigure}{.5\linewidth}
		\includegraphics[width=\linewidth]{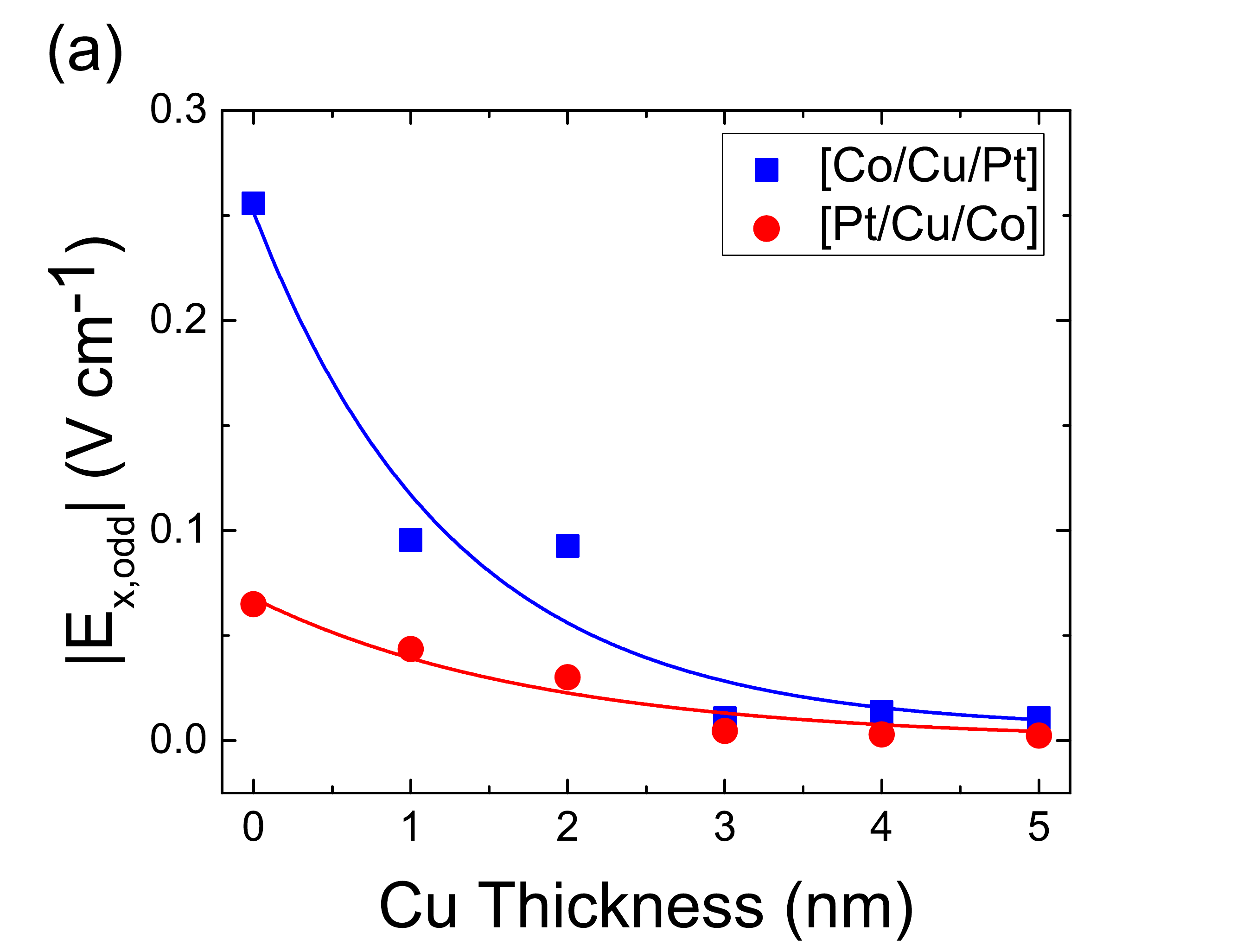}
	\end{subfigure}
	\begin{subfigure}{.5\linewidth}
		\includegraphics[width=\linewidth]{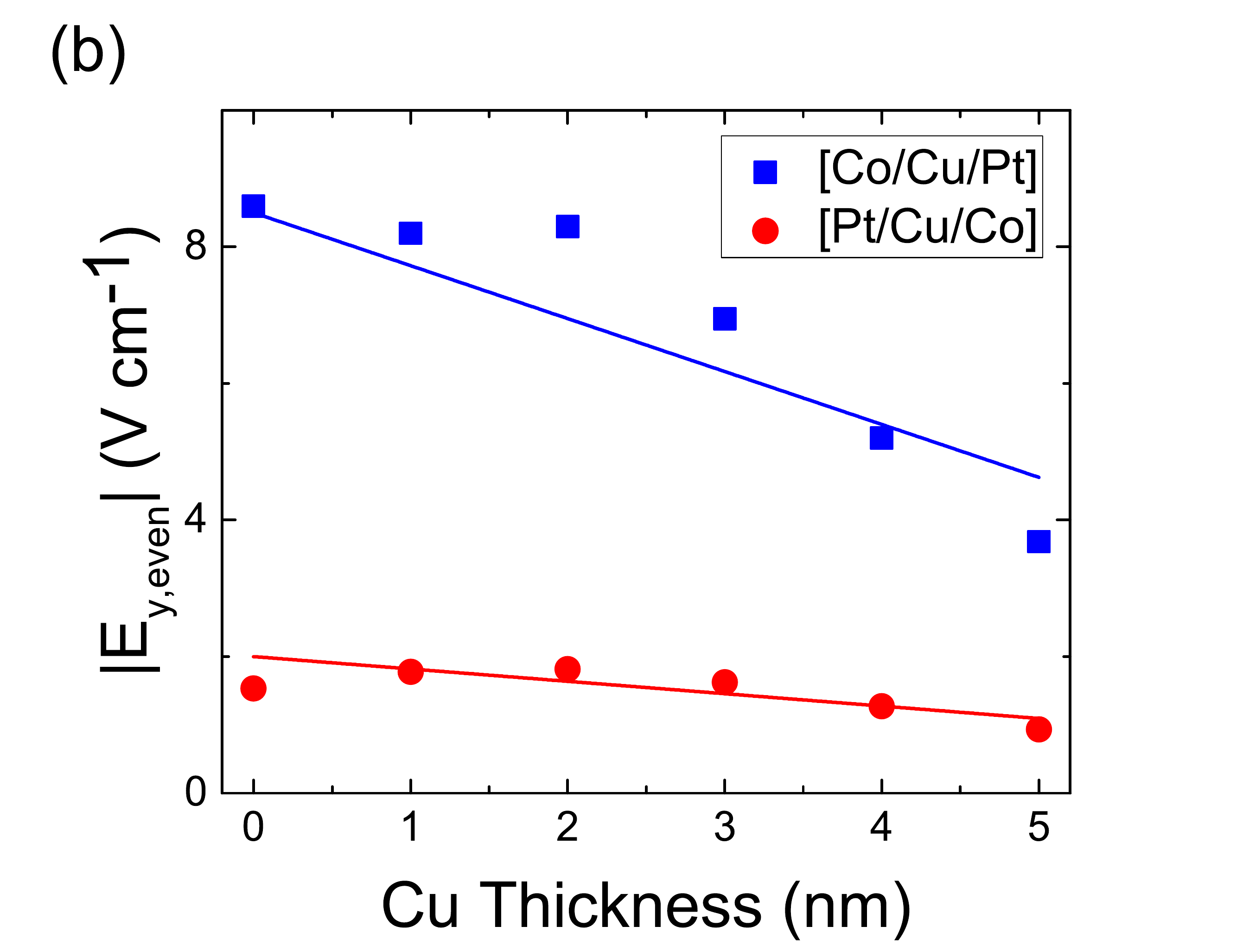}
	\end{subfigure}
	\caption{\label{fig:Cu_results} The integral strength of the THz electric field as a function of the Cu interlayer is obtained by integrating the amplitude of the Fourier spectrum $|E|$ over the frequency range between two cut-off frequencies. The cut-off frequencies were introduced as those corresponding to the half-maximum. (a) The helicity dependent $|E_{x,odd}|$ and (b) the helicity independent $|E_{y,even}|$ integral THz field. Circles (red) and squares (blue) correspond to the THz emission excited by a pump pulse with a central wavelength of 800 nm and a fluence of 0.6 $\mathrm{mJ}$ $\mathrm{cm^{-2}}$ incident on Co/Cu/Pt and Pt/Cu/Co, respectively. The solid lines are the fitted exponential decay functions.}
	\end{figure}
It is expected that laser excited electrons will be able to flow almost freely through the Cu interlayer, so one can anticipate qualitatively different results in the case of the Cu interlayer. We performed similar experiments on the Co/Cu/Pt samples using pump pulses with a central wavelength of 800 nm incident at Co/Cu/Pt and Pt/Cu/Co structures, respectively. We use the same analysis to extract the integrated field strength for the helicity dependent and independent fields. The thickness dependence of $|E_{x,odd}|$ and $|E_{y,even}|$ were fitted using Eq.\ref{eq:fit}. Figure \ref{fig:Cu_results} shows the thickness dependence of the helicity dependent and independent THz emission for the structures with the Cu interlayer. Both the helicity dependent and the helicity independent THz signals in Co/Cu/Pt scale differently with the interlayer thickness compared to the case of Co/ZnO/Pt. The THz electric field strength of the helicity dependent THz signal in Fig.\ref{fig:Cu_results}(a) shows an exponential decay compared to the immediate suppression in the case of ZnO. The integrated field strength of the helicity independent THz signal in Fig.\ref{fig:Cu_results}(b) show a nearly linear dependence on the Cu thickness compared to the exponential decay in the case of ZnO.

\section{Discussion}
\begin{table}
	\centering
	\begin{tabular}{l l l l}
		{\bf pump incident} & {\bf THz field} & $\boldsymbol{l}$ {\bf (nm)} & $\boldsymbol{\lambda_{pump}}$ {\bf (nm)} \\ \hline
		Co side & $E_{y,even}$ & 0.4 & 400\\
		Co side & $E_{y,even}$ & 0.4 & 800\\
		Pt side & $E_{y,even}$ & 0.3 & 800\\ \hline
	\end{tabular}
	\caption{\label{tab:zno_results} The decay lengths for the ZnO interlayer. The pump pulse is incident on either Co or Pt side of the sample (first column). We extract from the helicity dependent ($E_{x,odd}$) and independent ($E_{y,even}$) THz signal (second column) the decay lengths (third column). The samples were pumped with wavelengths of 400 nm and 800 nm (fourth columnn).}
\end{table}
\begin{table}
	\centering
	\begin{tabular}{l l l}
		{\bf pump incident} & {\bf THz field} & $\boldsymbol{l}$ {\bf (nm)} \\ \hline
		Co side & $E_{y,even}$ & 11 \\
		Pt side & $E_{y,even}$ & 11 \\
		Co side & $E_{x,odd}$  & 1.3  \\
		Pt side & $E_{x,odd}$  & 1.8  \\ \hline
	\end{tabular}
	\caption{\label{tab:cu_results} The decay lengths for the Cu interlayer. The pump pulse is incident on either Co or Pt side of the sample (first column). We extract from the helicity dependent ($E_{x,odd}$) and independent ($E_{y,even}$) THz signal (second column) the decay lengths (third column). The samples were all pumped with a wavelength of 800 nm.}
\end{table}
The experiments show that the introduction of an interlayer results in a decrease of the strength of the THz emission. Increasing the interlayer thickness leads to a further weakening of the THz emission. The decay rates deduced from the experiments are summarized in the Tables \ref{tab:zno_results} and \ref{tab:cu_results}. From the data three important conclusions can be drawn.

Firstly, the structures with the ZnO interlayer show about 28 times faster decay of the THz field strength of the helicity independent THz emission compared to the Cu interlayer structures. This fact shows that the conducting contact between Co and Pt layers is important for the mechanism of THz emission. In particular, this observation agrees with the expected microscopic mechanism in which THz radiation is generated as a result of a spin current from Co to Pt and spin to charge conversion of the current due to the inverse spin-Hall effect in Pt.

Secondly, it is seen that the decay length for the case of the pump being incident at Co/X/Pt is characterized by nearly the same decay length as for the case of the pump being incident at Pt/X/Co. Moreover, the THz field strength is weaker for the case when the pump is incident at Pt/X/Co (see Fig.\ref{fig:ZnO_odd-even} and \ref{fig:Cu_results}), which is due to absorption of the THz radiation in the glass substrate. Note that the electrons traveling from Pt to Co do not contribute to the discussed THz emission. If the light pulse first hits the other layers before it reaches Co, the light energy transfered to the electrons in Co is decreased due to the absorption in Pt, Cu, and ZnO.

Thirdly, the experiments in which Co/ZnO/Pt heterostructures were pumped with light at the wavelengths of 400 nm and 800 nm, respectively, also showed nearly the same decay lengths. This observation is again in agreement with the mechanism of generation of THz radiation due to the spin currents and inverse spin-Hall effect. Since the spin currents originate from ultrafast laser-induced demagnetization of Co, it is reasonable to expect that the strength of the spin currents and the demagnetization share the same dependence on the wavelength of the laser excitation. The efficiency and the speed of the demagnetization are hardly sensitive to the excitation wavelengths. Moreover, the THz spectrum does not change shape and its amplitude scales linearly with the pump fluence up to 1 $\mathrm{mJ}$ $\mathrm{cm^{-2}}$ \cite{huisman2016}. As a result, the demagnetization in the thermodynamical three-temperature model is only defined by the optical energy pumped into the electron subsystem \cite{melnikov2011,choi2014}. This is still the most popular model in ultrafast magnetism. Consequently, also for the spin current a change of the wavelength of the pump light does not lead to significant changes.
	
Since spin transport is the origin of the THz emission in Co/X/Pt structures, it is interesting to compare the decay lengths obtained in our experiment with values in literature. From Tab.\ref{tab:zno_results} we obtain an average ultrafast spin diffusion length of $l^{uf}_{sf} = 0.4$ nm for ZnO. From literature we find a spin diffusion length of $l_{sf} = 6.2$ nm obtained in thermal equilibrium from magneto-resistance measurements at 200 K by Althammer et al.\cite{althammer2012}. Our measured decay length is shorter in more than an order of magnitude. We also want to point out that the spin diffusion length in Ref.\cite{althammer2012} was measured at 200 K. Moreover, the spin diffusion length increases with decreasing temperature. While the decay length of Co/ZnO/Pt was measured at room-temperature it is still much shorter than the spin diffusion length of Ref.\cite{althammer2012}, if extrapolated to 300 K.
	
In the case of the Cu interlayer, the decay in the helicity independent THz field strength in Fig.\ref{fig:Cu_results}(b) shows a nearly linear relation. It can be seen as the first order Taylor series of the exponential decay. The average ultrafast spin diffusion length for Cu (see Tab.\ref{tab:cu_results}) is $l^{uf}_{sf} = 11$ nm and is of the same order as the spin-diffusion length ($l_{sf} = 38$ nm) obtained from magneto-resistance measurements in multilayered Co/Cu thin wires by Voegeli et al. \cite{voegeli1995}. However our decay length is still 3 times lower. Comparing with other spin diffusion lengths in literature, we find a much larger disagreement \cite{jedema2001, bass2007}. The main difference is that in our experiment the spin current pulses are generated at an ultrashort timescale and reveal quantities which describe transport properties of strongly non-equilibrium electrons, while conventional transport measurements study spin diffusion in the vicinity of thermodynamic equilibrium. Moreover, it should be noted that the spin diffusion length in Cu is strongly correlated to the bulk transport \cite{voegeli1995}. In our experiment the interlayer transport was measured in thin nm-layers, while the conventional transport experiments use 25 nm thick or even thicker layers with nearly bulk properties. It was also mentioned that the spin diffusion length in Cu is strongly correlated to the bulk resistivity \cite{voegeli1995}.

The effect of a varying Cu interlayer was studied before by Wu et al. \cite{wu2017}. It was found that the THz signal gradually decreases with Cu thickness, corresponding with our observations. It was concluded that the bulk inverse spin-Hall effect is the dominant mechanism of THz electric field radiation since no abrupt change from interfacial effects was observed upon insertion of a Cu interlayer.

The helicity dependent $E_{x,odd}$ THz signal is heavily dependent on the properties of the interface between Co and Pt where the space inversion symmetry is broken. It is seen from the thickness dependence of the structures with the ZnO interlayer (see Fig.\ref{fig:ZnO_odd-even}(a) and Fig.\ref{fig:ZnO_spectrum-results}(a)) that the helicity dependent current is significantly suppressed already at 1 nm separation of the Co and Pt layers. Replacing ZnO with Cu results in a completely different behavior. The helicity dependent THz signal gradually decreases with Cu thickness. It follows an exponential function with the decay length of $l = 1.5$ nm. All the observations of THz emission from samples with Cu and ZnO interlayers show that conductivity between the layers of Co and Pt is of crucial importance for the generation of the photocurrents.

\section{Conclusions}
Experimental studies of the THz emission induced by femtosecond laser pulses in Co/ZnO/Pt and Co/Cu/Pt multilayers reveal that the most efficient emission is observed for cases when electric conductivity between Pt and Co layers is high. Helicity independent laser-induced THz emission appears to be far less sensitive to the interlayer than the helicity dependent one. For instance, the helicity independent THz emission is observed with up to 5 nm thick interlayer of Cu and ZnO. As the helicity independent THz emission is explained as a result of the laser excited spin current from Co to Pt, the decay of the emission upon an increase of the interlayer thickness reveals an ultrafast spin-diffusion length for the hot spin-polarized electrons in the interlayers. At the same time, the helicity dependent THz emission disappears at 1 nm interlayer of ZnO and experiences an attenuation of roughly a factor of 10 for the case of 5 nm thick Cu interlayer. The helicity dependent THz emission is explained in terms of spin-galvanic effects for which it is important to have electrons that feel the magnetization of Co and spin-orbit interaction of Pt at the same time. Consequently, helicity dependent THz emission probes the ability of the spin-polarized electrons of Co and Pt to intermix.

\section{Acknowledgements}
This work was supported by the Nederlandse Organisatie voor Wetenschappelijk Onderzoek (NWO), the European Union’s Horizon 2020 research and innovation program under the FET-Open Grant Agreement No.713481 (SPICE), the European Research Council ERC Grant agreement No.339813 (Exchange), the Ministry of Education and Science of the Russian Federation (Project No.14.Z50.31.0034), and Russian Science Foundation (16-12-10520). We want to thank Tonnie Toonen and Sergey Semin for all the technical support and Thomas Huisman for the help with measurements.

\bibliography{ref}
\bibliographystyle{unsrt}

\end{document}